\documentclass[12pt]{article}
\usepackage{amsmath,enumerate,amsfonts,amssymb,color,graphicx}

\def\RR{{\mathbb R}}

\def\Ric{{\rm Ric}}

\newtheorem{theorem}{Theorem}[section]

\newtheorem{question}[theorem]{Question}

\def\Riem{{\rm Riem}}
\def\Ric{{\rm Ric}}

\DeclareFontFamily{OT1}{rsfs}{}
\DeclareFontShape{OT1}{rsfs}{m}{n}{ <-7> rsfs5 <7-10> rsfs7 <10-> rsfs10}{}
\DeclareMathAlphabet{\mycal}{OT1}{rsfs}{m}{n}

\newcounter{marnote}

\begin{document}

\title{A generalized mass involving higher order symmetric function of the curvature tensor}
\author{YanYan Li \thanks{Mathematics Department, Rutgers University. Partially supported by NSF Grant DMS-1203961}~~and Luc Nguyen \thanks{Mathematics Department, Princeton University}}
\maketitle

\begin{abstract}
We define a generalized mass for asymptotically flat manifolds using some higher order symmetric function of the curvature tensor. This mass is non-negative when the manifold is locally conformally flat and the $\sigma_k$ curvature vanishes at infinity. In addition, with the above assumptions, if the mass is zero, then, near infinity, the manifold is isometric to a Euclidean end.
\end{abstract}
\section{Introduction}

Let $(M,g)$ be an $n$-dimensional Riemannian manifold. Assume that $(M,g)$ is asymptotically flat, i.e. there is a compact set $K \subset M$, some $R \geq 1$ and a diffeomorphism $\Phi: M\setminus K  \rightarrow \RR^n \setminus B_R$ such that
\[
(\Phi^* g)_{ij}(x) = \delta_{ij} + o(|x|^{-\tau}) \text{ as } |x| \rightarrow \infty.
\]
If $\tau \geq \frac{n-2}{2}$ and the scalar curvature $R_g$ is integrable, then the so-called ADM mass of $(M,g)$ is defined by (see \cite{ADMMass,Bartnik86})
\[
m = \int_{S_\infty} (g_{ij,j} - g_{jj,i})dS^i,
\]
where $g_{ij,k}$ denotes a partial derivative and $dS^i$ is the normal surface element of $S_\infty$, the sphere at infinity.

That $m$ is a geometric invariant of $(M,g)$ is a consequence of the following expansion of the Hilbert-Einstein action:
\begin{equation}
R_g *1 = d(g^{ab}\,\omega_a{}^c \wedge \eta_{cb}) + g^{ab}\,\omega_d{}^c \wedge \omega_a{}^d \wedge \eta_{bc}
	\label{Eq:HEDecomp}
\end{equation}
where $R_g$ is the scalar curvature, $*1$ is the volume form, $\omega_a{}^b$ is the Levi-Civita connection one-form with respect to a frame $\{e_a\}$ and $\eta_{ab}= (e_a \wedge e_b) \rfloor *1$. See Bartnik \cite{Bartnik86} for more details.

Mass and its properties have attracted much attention since it was introduced. One of the reason is its wide range of applications in mathematical relativity and in geometric analysis. For example, consider the Yamabe problem which asks to find on a compact Riemannian manifold a conformal metric with constant scalar curvature. Its solutions are critical point of the Hilbert-Einstein functional in a fixed conformal class. Thus, it is not too surprising that the notion mass is useful in the study of the Yamabe problem. In fact, it is \emph{very important} in the solution of the Yamabe problem \cite{Schoen84} as well as in the resolution of compactness issue of the Yamabe problem \cite{Schoen91, LiZhu99, Druet04, LiZhang05, LiZhang07, Marques05, KMS09}.

In recent years, fully nonlinear versions of the Yamabe problem have received
much attention after the work of Viaclovsky \cite{Viac00-Duke, Viac00-TrAMS, Viac02} and of Chang, Gursky and Yang
\cite{CGY02-AnnM, CGY02-JAM, CGY03-IHES, CGY03-IP}; see e.g. \cite{ GW03-JRAM, LiLi03,  LiLi05, GeWang06, GV07, STW07,  TW09,  LiNg-arxiv,  TW10}. For a metric $g$, let $A_g$ be the Schouten tensor of $g$, i.e.
\[
A_g = \frac{1}{n-2}\Big(\Ric_g- \frac{1}{2(n-1)}\,R_g\,
g\Big), 
\]
where $\Ric_g$ and $R_g$ denote the Ricci curvature and the scalar curvature of $g$. Let $\lambda(A_g)=(\lambda_1, \cdots, \lambda_n)$
denote the eigenvalues of
$A_g$ with respect to $g$. For $1\leq k \leq n,$ let $\sigma_k(\lambda) = \sum_{1\leq i_1 <\cdots <
i_k\leq n}\lambda_{i_1}\cdots \lambda_{i_k}, \lambda = (\lambda_1,
\cdots, \lambda_n) \in \Bbb R^n$, denote the $k$-th elementary symmetric
function, and let $\Gamma_k$ denote the connected component of 
$\{\lambda \in \Bbb R^n | \sigma_k (\lambda) > 0\}$ containing the
positive cone $\{\lambda \in \Bbb R^n|\lambda_1, \cdots, \lambda_n >
0\}$.

\begin{question}\label{MQuest}
Let $(N,h)$ be a compact, smooth
Riemannian manifold of dimension $n\ge 3$ satisfying
$\lambda(A_h)\in \Gamma$ on $N$ and $1 \leq k \leq n$. Is there a
smooth positive function $u$ on $N$ such that
$\hat h= u^{\frac{4}{n-2}} h$ satisfies
\begin{equation}
\sigma_k(\lambda(A_{\hat h})) =1, \quad \lambda (A_{\hat g}) \in \Gamma_k,
\quad \mbox{ on } N?
\label{9}
\end{equation}
\end{question}

Equation (\ref{9}) is a second
order fully nonlinear elliptic equation of $u$.
The special case of Question \ref{MQuest} 
for  
$k = 1$ is the 
Yamabe problem in the so-called positive case.

The problem is in general not a variational one when $k \geq 3$. 
Natural variants of equation
(\ref{9}) 
which are of variational form
  have been introduced 
by Chang and Fang \cite{ChangFang08}; see also
a subsequent paper of Graham \cite{Graham09}
on the algebraic structure  of these equations under
conformal transformations.

From the discussion on the relation between the ADM mass and the Yamabe problem, it is natural to ask if there is some notion of mass associated with the $\sigma_k$ Yamabe problem. The main goal of this note is to give some generalization along this line. While we have not been able to identify such a notion that is directly related to the $\sigma_k$ curvature, we are able to do so for a variant for $2 \leq k < \frac{n}{2}$, which coincides with the $\sigma_k$ curvature when $(M,g)$ is locally conformally flat. This is motivated by a relation between the Pffafian and the $\sigma_{n/2}$ curvature (see Viaclovsky \cite{Viac00-Duke}). See Theorem \ref{Thm:MassDef} for a precise definition.

After our work was done, the Yanyan Li heard a talk by Guofang Wang in the conference on `Geometric PDEs' in the trimester on `Conformal and K\"ahler Geometry' at IHP in which he announced that he together with Yuxin Ge and Jie Wu had also developed a notion of higher order mass. After the talk, Yanyan informed Guofang that we had
defined a mass by using an invariant of the $\sigma_k$
curvature  which agrees with the $\sigma_k$ curvature when
the manifold
is locally conformally flat and proved that the mass is
non-negative
under the assumption that the manifold is locally
conformally flat and the
$\sigma_k$ curvature is zero near infinity, together with a
  rigidity in that case.

The rest of the paper is organized as follows. In Section \ref{Sec:CurvInv}, we define a curvature invariant $\Lambda_k$ which coincides with the $\sigma_k$ curvature when $(M,g)$ is locally conformally flat. We also provide a decomposition for $\Lambda_k$ which is a generalization of the decomposition \eqref{Eq:HEDecomp} for the Hilbert-Einstein action. In Section \ref{Sec:MassDef}, we used the decomposition developed in Section \ref{Sec:CurvInv} to define a mass, called the $k$-th mass. In Section \ref{Sec:PosMass}, we announce a very restrictive version of the positive mass theorem for the $k$-th mass.

\bigskip
\noindent{\bf Acknowledgment.} The authors would like to thank Professor Deser for his remarks which help improve the presentation of the paper.

\section{A curvature invariant} \label{Sec:CurvInv}

Consider an $n$-dimensional Riemannian manifold $(M,g)$. Let $\{e_1, \ldots, e_n\}$ be a local orthonormal frame and $\{\theta^1, \ldots, \theta^n\}$ its dual coframe. Let $(\omega_i{}^j)$ be the skew-symmetric matrix of Levi-Civita connection one-forms:
\[
\nabla_\xi e_i = \omega_i{}^j(\xi)\,e_j.
\]
(Here and below, upper indices label rows while lower indices label columns.) The first structural equations read
\[
d\theta^j = \theta^i \wedge \omega_i{}^j.
\]
The curvature tensor is viewed as a skew-symmetric matrix of two-forms,
\[
\Omega_i{}^j(X,Y) = \theta^j(R(X,Y)e_i).
\]
The second structural equations read
\[
\Omega_i{}^j = d\omega_i{}^j - \omega_i{}^k \wedge \omega_k{}^j.
\]
Also, the first and second Bianchi identities read
\[
\theta^i \wedge \Omega_i{}^j = 0 \text { and } d\Omega_i{}^j = -\Omega_i{}^k \wedge \omega_k{}^j + \omega_i{}^k \wedge \Omega_k{}^j.
\]

For a multi-index $I = (i_1, \ldots, i_m)$, let $\theta^I = \theta^{i_1} \wedge \ldots \wedge \theta^{i_m}$ and $\theta^{[I]} = *\theta^{I}$ where $*$ is the Hodge dual operator. 

For $1 \leq k \leq \frac{n}{2}$, define the $n$-form
\[
\Lambda_k = \frac{1}{2^k\,k!}\sum_{I = (i_1, \ldots, i_{2k}) \in S_{n,2k}} \Omega_{i_1}{}^{i_2} \wedge \ldots \wedge \Omega_{i_{2k-1}}{}^{i_{2k}} \wedge \theta^{[I]},
\]
where 
\[
S_{n,2k} := \Big\{(i_1, \ldots, i_{2k}): 1 \leq i_p \leq n, i_p \neq i_q \text{ whenever } p \neq q\Big\}.
\]
It is useful to observe that
\[
\Lambda_k = \frac{1}{2^k\,k!}\sum_{1 \leq i_1, \ldots, i_{2k} \leq n} \Omega_{i_1}{}^{i_2} \wedge \ldots \wedge \Omega_{i_{2k-1}}{}^{i_{2k}} \wedge \theta^{[I]}.
\]

For $k = 1$, $*\Lambda_1$ is half the scalar curvature. For $n$ even and $k = n/2$, $\Lambda_{n/2}$ is the Pfaffian. Those quantities are frame independent. We claim that this is true for all $\Lambda_k$'s. Indeed, let $P$ be an orthonormal matrix function, $\tilde e_i = P_i{}^j\,e_j$, $\tilde\theta^j = (P^{-1})_i{}^j\,\theta^i$. We have
\[
\tilde\omega_i{}^j = P_i{}^k\,\omega_k{}^l\,(P^{-1})_l{}^j + P_i^{k}\,d(P^{-1})_k{}^j \text{ and } \tilde\Omega_i{}^j = P_i{}^k\,\Omega_k{}^l\,(P^{-1})_l{}^j.
\]
We then have
\begin{align*}
2^k\,k!\tilde\Lambda_k 
	&= \sum_{1 \leq i_1, \ldots, i_{2k} \leq n} \tilde\Omega_{i_1}{}^{i_2} \wedge \ldots \wedge \tilde\Omega_{i_{2k-1}}{}^{i_{2k}} \wedge \tilde\theta^{[I]}\\
	&= \sum_{1 \leq i_1, \ldots, i_{2k} \leq n} \big[P_{i_1}{}^{p_1}\,(P^{-1})_{p_2}{}^{i_2} \Omega_{p_1}{}^{p_2}\big] \wedge \ldots \wedge \big[ P_{i_{2k-1}}{}^{p_{2k-1}}\,(P^{-1})_{p_{2k}}{}^{i_{2k}} \Omega_{p_{2k-1}}{}^{p_{2k}}\big]\\
		&\qquad\qquad \wedge *\Big(\big[(P^{-1})_{q_1}{}^{i_1}\,\theta^{q_1}\big] \wedge \ldots \wedge \big[(P^{-1})_{q_{2k}}{}^{i_{2k}}\,\theta^{q_{2k}}\big]\Big)\\
	&= \sum_{1 \leq i_1, \ldots, i_{2k} \leq n} P_{i_1}{}^{p_1}\,(P^{-1})_{q_1}{}^{i_1}\,(P^{-1})_{p_2}{}^{i_2}\,(P^{-1})_{q_2}{}^{i_2}\ldots \\
		&\qquad\qquad \ldots P^{-1}_{i_{2k-1}}{}^{p_{2k-1}}\,(P^{-1})_{q_{2k-1}}{}^{i_{2k-1}}\,(P^{-1})_{p_{2k}}{}^{i_{2k}}\,(P^{-1})_{q_{2k}}{}^{i_{2k}}\\
		&\qquad \qquad \Omega_{p_1}{}^{p_2} \wedge \ldots \wedge \Omega_{p_{2k-1}}{}^{p_{2k}} \wedge *(\theta^{q_1} \wedge \ldots \wedge \theta^{q_{2k}})\\
	&= \sum_{1 \leq p_1, \ldots, p_{2k} \leq n} \Omega_{p_1}{}^{p_2} \wedge \ldots \wedge \Omega_{p_{2k-1}}{}^{p_{2k}} \wedge *(\theta^{p_1} \wedge \ldots \wedge \theta^{p_{2k}})\\
	&= 2^k\,k!\Lambda_k,
\end{align*}
where in the second-to-last identity we have used the orthogonality of $P$.

When $g$ is conformally flat, $*\Lambda_k$ is proportional to the $\sigma_k$-curvature. This was noticed by Viaclovsky \cite{Viac00-Duke} in case $k = n/2$. The argument for general $k$ is similar. We include it here for completeness. Recall that the Riemann curvature tensor $\Riem$ admits the decomposition
\[
\Riem = W_g + A_g \odot g
\]
where $\odot$ is the Kulkarni-Nomizu product and $W_g$ is the Weyl tensor of $g$. When $(M,g)$ is locally conformally flat, $W_g \equiv 0$. Fix a point $p \in M$. The local orthonormal frame $\{e_1, \ldots, e_n\}$ is chosen so that $A_g$ is diagonalized at $p$ with eigenvalue $\lambda_1, \ldots, \lambda_n$; in particular, $A_i{}^j = \lambda_i\,\delta_{ij}$. We have at $p$ that
\begin{align*}
\Omega_i{}^j 
	&= \Riem_i{}^j{}_{kl}\,\theta^k\wedge \theta^l\\
	&= (A_g \odot g)_i{}^j{}_{kl}\,\theta^k\wedge \theta^l\\
	&= (A_{ik}\,\delta_{jl} - A_{jk}\,\delta_{il} + A_{jl}\,\delta_{ik} - A_{il}\,\delta_{jk})\,\theta^k\wedge \theta^l\\
	&= 2(\lambda_i + \lambda_j)\,\theta^i \wedge \theta^j.
\end{align*}
It follows that, also at $p$,
\begin{align*}
\Lambda_k 
	&= \frac{1}{2^k\,k!}\sum_{I = (i_1, \ldots, i_{2k}) \in S_{n,2k}} \Omega_{i_1}{}^{i_2} \wedge \ldots \wedge \Omega_{i_{2k-1}}{}^{i_{2k-2}} \wedge \theta^{[I]}\\
	&= \frac{1}{k!} \sum_{I = (i_1, \ldots, i_{2k}) \in S_{n,2k}} (\lambda_{i_1} + \lambda_{i_2})\ldots(\lambda_{i_{2k-1}} + i_{2k}) \underbrace{\theta^I  \wedge \theta^{[I]}}_{=dv_g}\\
	&= \frac{1}{k!} \sum_{I = (i_1, \ldots, i_{2k}) \in S_{n,2k}} (\lambda_{i_1} + \lambda_{i_2})\ldots(\lambda_{i_{2k-1}} + i_{2k}) dv_g\\
	&= \frac{2^{k}(n-k)!}{k!(n-2k)!} \sum_{J = (j_1, \ldots, j_k) \in S_{n,k}} \lambda_{j_1}\ldots \lambda_{j_k}\, dv_g\\
	&= \frac{2^k(n-k)!}{(n-2k)!} \sigma_k(A_g)\,dv_g,
\end{align*}
i.e. $\Lambda_k$ is proportional to $\sigma_k(A_g)\,dv_g$.

To finish this section, we derive a decomposition of $\Lambda_k$ which we will need later. By the second structural equations,
\begin{align*}
2^k\,k!\Lambda_k 
	&= \sum_{1 \leq i_1, \ldots, i_{2k} \leq n} \Omega_{i_1}{}^{i_2} \wedge \ldots \wedge \Omega_{i_{2k-3}}{}^{i_{2k-2}} \wedge d\omega_{i_{2k-1}}{}^{i_{2k}} \wedge \theta^{[I]}\\
		&\qquad - \sum_{1 \leq i_1, \ldots, i_{2k} \leq n} d\Omega_{i_1}{}^{i_2} \wedge \ldots \wedge \Omega_{i_{2k-3}}{}^{i_{2k-2}} \wedge \omega_{i_{2k-1}}{}^{p_{k}} \wedge \omega_{p_k}{}^{i_{2k}} \wedge \theta^{[I]}\\
	&= \sum_{1 \leq i_1, \ldots, i_{2k} \leq n} d\Big(\Omega_{i_1}{}^{i_2} \wedge \ldots \wedge \Omega_{i_{2k-3}}{}^{i_{2k-2}} \wedge \omega_{i_{2k-1}}{}^{i_{2k}} \wedge \theta^{[I]}\Big)\\
		&\qquad - \sum_{1 \leq i_1, \ldots, i_{2k} \leq n} d\Big(\Omega_{i_1}{}^{i_2} \wedge \ldots \wedge \Omega_{i_{2k-3}}{}^{i_{2k-2}}\Big) \wedge \omega_{i_{2k-1}}{}^{i_{2k}} \wedge \theta^{[I]}\\
		&\qquad + \sum_{1 \leq i_1, \ldots, i_{2k} \leq n} \Omega_{i_1}{}^{i_2} \wedge \ldots \wedge \Omega_{i_{2k-3}}{}^{i_{2k-2}} \wedge \omega_{i_{2k-1}}{}^{i_{2k}} \wedge d\theta^{[I]}\\
		&\qquad - \sum_{1 \leq i_1, \ldots, i_{2k} \leq n} \Omega_{i_1}{}^{i_2} \wedge \ldots \wedge \Omega_{i_{2k-3}}{}^{i_{2k-2}} \wedge \omega_{i_{2k-1}}{}^{p_{k}} \wedge \omega_{p_k}{}^{i_{2k}} \wedge \theta^{[I]}.
\end{align*}

Note that, by symmetry,
\begin{align*}
&\sum_{1 \leq i_1, \ldots, i_{2k} \leq n} d\Omega_{i_1}{}^{i_2} \wedge \Omega_{i_3}{}^{i_4} \ldots \wedge  \Omega_{i_{2k-3}}{}^{i_{2k-2}} \wedge  \omega_{i_{2k-1}}{}^{i_{2k}} \wedge \theta^{[I]}\\
	&\qquad = \ldots = \sum_{1 \leq i_1, \ldots, i_{2k} \leq n} \Omega_{i_1}{}^{i_2} \wedge \ldots\wedge \Omega_{i_{2k-5}}{}^{i_{2k-4}}  \wedge d\Omega_{i_{2k-3}}{}^{i_{2k-2}} \wedge  \omega_{i_{2k-1}}{}^{i_{2k}} \wedge \theta^{[I]},
\end{align*}
which, by the second Bianchi identity and anti-symmetry, is equal to
\begin{align*}
&\sum_{1 \leq i_1, \ldots, i_{2k} \leq n} \Omega_{i_1}{}^{i_2} \wedge \ldots \wedge \Omega_{i_{2k-5}}{}^{i_{2k-4}} \wedge\\
			&\qquad\qquad \wedge (-\Omega_{i_{2k-3}}{}^{r_{k-1}}\wedge \omega_{r_{k-1}}{}^{i_{2k-2}} + \omega_{i_{2k-3}}{}^{r_{k-1}} \wedge \Omega_{r_{k-1}}{}^{i_{2k-2}}) \wedge  \omega_{i_{2k-1}}{}^{i_{2k}} \wedge \theta^{[I]}\\
&= -2\sum_{1 \leq i_1, \ldots, i_{2k} \leq n} \Omega_{i_1}{}^{i_2} \wedge \ldots \wedge \Omega_{i_{2k-5}}{}^{i_{2k-4}} \wedge \Omega_{i_{2k-3}}{}^{r_{k-1}} \wedge \omega_{r_{k-1}}{}^{i_{2k-2}}  \wedge  \omega_{i_{2k-1}}{}^{i_{2k}} \wedge \theta^{[I]}.
\end{align*}

Next, we compute $d\theta^{[I]}$. Assume first that $I = (i_1, \ldots, i_{2k}) \in S_{n,2k}$. We supplement $I$ with $i_{2k+1}, \ldots, i_n$ so that $(i_1, \ldots, i_n)$ is a permutation of $(1,\ldots, n)$. We have
\[
\theta^{[I]} =  \delta_{1 \ldots n}^{i_1\ldots i_n} \theta^{i_{2k+1}} \wedge \ldots \wedge \theta^{i_n}.
\]
In view of the first structural equations, this implies that
\begin{align*}
d\theta^{[I]} 
	&= \delta_{1 \ldots n}^{i_1\ldots i_n} \sum_{s = 2k+1}^n \sum_{t = 1}^{2k} (-1)^{s - 1}\theta^{i_{2k+1}} \wedge \ldots \wedge\theta^{i_{s - 1}} \wedge (\theta^{i_t} \wedge \omega_{i_t}{}^{i_{s}}) \wedge \theta^{i_{s+1}} \wedge \ldots \wedge \theta^{i_n}\\
	&= \delta_{1 \ldots n}^{i_1\ldots i_n}\sum_{t = 1}^{2k} \sum_{s = 2k+1}^n (-1)^s \omega_{i_t}{}^{i_{s}} \wedge  \theta^{i_t} \wedge \theta^{i_{2k+1}} \wedge \ldots \wedge\theta^{i_{s - 1}} \wedge  \theta^{i_{s+1}} \wedge \ldots \wedge \theta^{i_n}\\
	&= \sum_{t = 1}^{2k} \sum_{s = 2k+1}^n \omega_{i_t}{}^{i_{s}} \wedge \theta^{[I: i_t \rightarrow i_s]}\\
	&= \sum_{t = 1}^{2k} \sum_{s = 1}^n \omega_{i_t}{}^{s} \wedge \theta^{[I: i_t \rightarrow s]},
\end{align*}
where $I: i_t \rightarrow i_s$ denotes $(i_1, \ldots, i_{t-1}, i_s, i_{t+1}, \ldots, i_{2k})$. This continues to holds when for general multi-index $I = (i_1, \ldots, i_{2k}) \in \{1, \ldots, n\}^k$. 
Indeed, if $I \notin S_{n,2k}$, we have $d\theta^{[I]} = 0$ and
\begin{align*}
&\sum_{t = 1}^{2k} \sum_{s = 1}^n \omega_{i_t}{}^{s} \wedge \theta^{[I: i_t \rightarrow s]}\\
	&\qquad = \sum_{\text{special }t's} \sum_{s = 1}^n \omega_{i_t}{}^{s} \wedge \theta^{[I: i_t \rightarrow s]}\\
	&\qquad = \frac{1}{2}\sum_{\text{special }t's} \sum_{s = 1}^n (\omega_{i_t}{}^{s} \wedge \theta^{[I: i_t \rightarrow s]} + \omega_{i_{\tilde t}}{}^{s} \wedge \theta^{[I: i_{\tilde t} \rightarrow s]})\\
	&\qquad = 0.
\end{align*}
where the set of special $t$'s are those such that there is a unique $\tilde t \neq t$ such that $i_t = i_{\tilde t}$. It thus follows that
\begin{align*}
&\sum_{1 \leq i_1, \ldots, i_{2k} \leq n} \Omega_{i_1}{}^{i_2} \wedge \ldots \wedge \Omega_{i_{2k-3}}{}^{i_{2k-2}} \wedge \omega_{i_{2k-1}}{}^{i_{2k}} \wedge d\theta^{[I]}\\
	&\qquad = \sum_{1 \leq i_1, \ldots, i_{2k} \leq n} \Omega_{i_1}{}^{i_2} \wedge \ldots \wedge \Omega_{i_{2k-3}}{}^{i_{2k-2}} \wedge \omega_{i_{2k-1}}{}^{i_{2k}} \wedge \sum_{t = 1}^{2k}\sum_{s = 1}^n \omega_{i_t}{}^{s} \wedge \theta^{[I: i_t \rightarrow s]}\\
	&\qquad = 2(k-1)\sum_{1 \leq i_1, \ldots, i_{2k}, s \leq n} \Omega_{i_1}{}^{i_2} \wedge \ldots \wedge \Omega_{i_{2k-3}}{}^{i_{2k-2}} \wedge \omega_{i_{2k-1}}{}^{i_{2k}} \wedge \omega_{i_{2k-2}}{}^{s} \wedge \theta^{[I: i_{2k-2} \rightarrow s]}\\
		&\qquad\qquad + 2\sum_{1 \leq i_1, \ldots, i_{2k},s \leq n} \Omega_{i_1}{}^{i_2} \wedge \ldots \wedge \Omega_{i_{2k-3}}{}^{i_{2k-2}} \wedge \omega_{i_{2k-1}}{}^{i_{2k}} \wedge  \omega_{i_{2k}}{}^{s} \wedge \theta^{[I: i_{2k} \rightarrow s]}\\
	&\qquad = -2(k-1)\sum_{1 \leq j_1, \ldots, j_{2k},i_{2k-2} \leq n} \Omega_{j_1}{}^{j_2} \wedge \ldots \wedge \Omega_{j_{2k-3}}{}^{i_{2k-2}} \wedge \omega_{i_{2k-2}}{}^{j_{2k-2}} \wedge \omega_{j_{2k-1}}{}^{j_{2k}} \wedge \theta^{[J]}\\
		&\qquad\qquad + 2\sum_{1 \leq j_1, \ldots, j_{2k}, i_{2k} \leq n}  \Omega_{j_1}{}^{j_2} \wedge \ldots \wedge \Omega_{j_{2k-3}}{}^{j_{2k-2}} \wedge \omega_{j_{2k-1}}{}^{i_{2k}} \wedge \omega_{i_{2k}}{}^{j_{2k}} \wedge  \theta^{[J]}.
\end{align*}

We thus get
\begin{align*}
2^k\,k!\Lambda_k 
	&= \sum_{1 \leq i_1, \ldots, i_{2k} \leq n} d\Big(\Omega_{i_1}{}^{i_2} \wedge \ldots \wedge  \Omega_{i_{2k-3}}{}^{i_{2k-2}} \wedge  \omega_{i_{2k-1}}{}^{i_{2k}} \wedge \theta^{[I]}\Big)\\
		&\qquad + \sum_{1 \leq i_1, \ldots, i_{2k}, p_k \leq n} \Omega_{i_1}{}^{i_2} \wedge \ldots \wedge \Omega_{i_{2k-3}}{}^{i_{2k-2}} \wedge \omega_{i_{2k-1}}{}^{p_{k}} \wedge \omega_{p_k}{}^{i_{2k}} \wedge \theta^{[I]}.
\end{align*}

Set
\begin{align*}
\Lambda^1_k 
	&= \frac{1}{2^k\,k!}\sum_{1 \leq i_1, \ldots, i_{2k} \leq n} \Omega_{i_1}{}^{i_2} \wedge \ldots \wedge  \Omega_{i_{2k-3}}{}^{i_{2k-2}} \wedge  \omega_{i_{2k-1}}{}^{i_{2k}} \wedge \theta^{[I]},\\
\Lambda^2_k
	&= \frac{1}{2^k\,k!}\sum_{1 \leq i_1, \ldots, i_{2k}, p_k \leq n} \Omega_{i_1}{}^{i_2} \wedge \ldots \wedge \Omega_{i_{2k-3}}{}^{i_{2k-2}} \wedge \omega_{i_{2k-1}}{}^{p_{k}} \wedge \omega_{p_k}{}^{i_{2k}} \wedge \theta^{[I]}.
\end{align*}
Then $\Lambda_k = d\Lambda^1_k + \Lambda^2_k$. It should be noted that, unlike $\Lambda_k$, $\Lambda^1_k$ and $\Lambda^2_k$ are frame dependent.

\section{Higher order mass}\label{Sec:MassDef}

Let $(M^n,g)$ be a Riemannian manifold and assume that there is a compact set $K \subset M$ such that $M\setminus K$ has an asymptotic flat structure of order $\tau$: There are some $R \geq 1$ and a diffeomorphism $\Phi: M\setminus K  \rightarrow \RR^n \setminus B_R$ such that
\[
(\Phi^* g)_{ij}(x) = \delta_{ij} + o_2(|x|^{-\tau}) \text{ as } |x| \rightarrow \infty,
\]
where $x = (x^1, \ldots, x^n)$ is the coordinate function with respect to $\Phi$  and we write $f = o_l(|x|^{-\tau})$ if $\partial_{i_1} \ldots \partial_{i_p} f = o(|x|^{-\tau - p})$ for any $1 \leq p \leq l$. We also assume that 
\[
\Lambda_k \in L^1(M).
\]
For simplicity, we only consider the case where $M$ has one end; the general case requires minor modification.

With respect to the asymptotic structure $\Phi$, let $e_i = \partial_{x^i}$ and $\theta^i = dx^i$. The connection one-forms and curvature two-forms are defined by
\[
\nabla_X e_i = \omega_i{}^j(X)\,e_j \text{ and } \Omega_i{}^j(X,Y) = \theta^j(R(X,Y)e_i).
\]
We note that $\omega_i{}^j$ and $\Omega_i{}^j$ may not be skew-symmetric. Define the $(n-1)$-form
\[
\Lambda^{1}_k(\Phi) = \frac{1}{2^k\,k!}\sum_{1 \leq i_1, \ldots, i_{2k} \leq n} \Omega_{i_1}{}^{i_2} \wedge \ldots \wedge  \Omega_{i_{2k-3}}{}^{i_{2k-2}} \wedge  \omega_{i_{2k-1}}{}^{i_{2k}} \wedge \theta^{[I]}
\]
and the associate ``$k$-th order mass''
\[
m_k(\Phi) = \lim_{R \rightarrow \infty} \int_{S_R} (-1)^k\,\Lambda^1_k(\Phi).
\]
Here $S_R$ denotes the coordinate sphere of radius $R$ centered at the origin. 

To show that $m_k(\Phi)$ is well-defined, we use the decomposition of $\Lambda_k$ which we derived earlier (in an orthonormal frame). We first use the Gram-Schmidt orthogonalization to construct an orthonormal frame:
\begin{align*}
\tilde e_1 
	&= \frac{e_1}{|e_1|},\\
\tilde e_2 
	&= \frac{e_2 - g(e_2,\tilde e_1)\,\tilde e_1}{|e_2 - g(e_2,\tilde e_1)\,\tilde e_1|},\\
\ldots\\
\tilde e_n
	&= \frac{e_n - g(e_n,\tilde e_1)\,\tilde e_1 - \ldots - g(e_n, \tilde e_{n-1})\tilde e_{n-1}}{|e_n - g(e_n,\tilde e_1)\,\tilde e_1 - \ldots - g(e_n, \tilde e_{n-1})\tilde e_{n-1}|}.
\end{align*}
Evidently,
\[
\tilde e_i = e_i + o_2(|x|^{-\tau}).
\]
Let $\{\tilde\theta^i\}$ be the dual frame to $\{\tilde e_i\}$ and define the connection one-forms $\tilde \omega_i{}^j$ and curvature two-forms $\tilde \Omega_i{}^j$ accordingly. Set
\begin{align*}
\tilde\Lambda^1_k 
	&= \frac{1}{2^k\,k!}\sum_{1 \leq i_1, \ldots, i_{2k} \leq n} \tilde\Omega_{i_1}{}^{i_2} \wedge \ldots \wedge \tilde \Omega_{i_{2k-3}}{}^{i_{2k-2}} \wedge  \tilde\omega_{i_{2k-1}}{}^{i_{2k}} \wedge \tilde\theta^{[I]},\\
\tilde\Lambda^2_k
	&= \frac{1}{2^k\,k!}\sum_{1 \leq i_1, \ldots, i_{2k}, p_k \leq n} \tilde\Omega_{i_1}{}^{i_2} \wedge \ldots \wedge \tilde\Omega_{i_{2k-3}}{}^{i_{2k-2}} \wedge \tilde\omega_{i_{2k-1}}{}^{p_{k}} \wedge \tilde\omega_{p_k}{}^{i_{2k}} \wedge \tilde\theta^{[I]}.
\end{align*}
Then $\Lambda_k = d\tilde\Lambda^1_k + \tilde\Lambda^2_k$.

To relate $\tilde \omega_i{}^j$ and $\tilde \Omega_i{}^j$ to $\omega_i{}^j$ and $\Omega_i{}^j$, we write 
\[
\tilde e_i = P_i{}^j\,e_j \text{ and } \tilde\theta^j = (P^t)_i{}^j\,\theta^i
\]
where the matrix $P$ satisfies
\[
P_i{}^j = \delta_{ij} + o_2(|x|^{-\tau}).
\]

We have
\[
\nabla_X \tilde e_i = \nabla_X (P_i{}^j\,e_j) = (dP_i{}^j(X) + P_i{}^k\,\omega_k{}^j(X))e_j = (dP_i{}^j(X) + P_i{}^k\,\omega_k{}^j(X))(P^{-1})_j{}^l\,\tilde e_l,
\]
which implies that
\[
\tilde\omega_i{}^j = \omega_i{}^j + dP_i{}^j + o_1(|x|^{-2\tau - 1}).
\]
Likewise
\[
\tilde \Omega_i{}^j = \tilde \theta_j(R(X,Y)\tilde e_i) = (P^t)_j{}^k P_i{}^l\,\theta_k(R(X,Y)e_l) = (P^t)_j{}^k P_i{}^l\,\Omega_k{}^l = \Omega_i{}^j + o(|x|^{-2\tau - 2}).
\]

From the above computation, we see that 
\[
\tilde\Lambda^2_k = O(|x|^{-(\tau(k+1) + 2k)}) \in L^1(M),
\]
and
\begin{align*}
\tilde \Lambda^1_k - \Lambda^1_k
	&= d\Big(\frac{1}{2^k\,k!}\sum_{1 \leq i_1, \ldots, i_{2k} \leq n} [P_{i_{2k-1}}{}^{i_{2k}} - \delta_{i_{2k-1}i_{2k}}] \Omega_{i_1}{}^{i_2} \wedge \ldots \wedge  \Omega_{i_{2k-3}}{}^{i_{2k-2}} \wedge \theta^{[I]}\Big)\\
		&\qquad+ o(|x|^{-(\tau(k+1) + 2k) + 1}).	
\end{align*}
These lead to, for $R \gg N \gg R_0$, 
\begin{align*}
\int_{S_R} \Lambda^1_k 
	&= \int_{S_R} \tilde\Lambda^1_k + o(R^{n - (\tau(k+1) + 2k)})\\
	&= \int_{S_{N}} \tilde\Lambda^1_k 
		+ \int_{B_R \setminus B_{N}} \Big[\Lambda_k + \tilde\Lambda^2_k\Big]
			+ o(R^{n - (\tau(k+1) + 2k)}).
\end{align*}
Hence, for
\[
\tau > \frac{n-2k}{k+1}
\]
the mass $m_k(\Phi)$ is well-defined.

In fact, the argument above shows that if $D_j$ is an exhaustion of $M$ by closed sets such that 
\[
R_j = \inf\{ |x|: x \in \partial D_k\} \rightarrow \infty \text{ and } R_j^{-(n-1)}|\partial D_k| \text{ remains bounded as } j \rightarrow \infty
\]
then
\[
m_k(\Phi) = \lim_{j \rightarrow \infty } \int_{S_j} (-1)^k\,\Lambda^1_k.
\]

We show next that $m_k(\Phi)$ is independent of the asymptotic structure $\Phi$. Assume that $\tilde \Phi$ is another asymptotic structure of $(M,g)$ and let $\tilde x = (\tilde x^1, \ldots, \tilde x^n)$ denotes the coordinate function with respect to $\tilde \Phi$. To show that $m_k(\Phi) = m_k(\tilde \Phi)$ we appeal to a theorem of Bartnik \cite[Theorem 3.1]{Bartnik86} to find harmonic coordinates $y = (y^1, \ldots, y^n)$ and $\tilde y = (\tilde y^1, \ldots, \tilde y^n)$ such that
\begin{align*}
&|x^i - y^i| + |\tilde x^i - \tilde y^i| = o(|x|^{1 - \tau}) = o(|\tilde x|^{1 - \tau}),\\
&|g(\partial_{x^i}, \partial_{x^j}) - g(\partial_{y^i}, \partial_{y^j})| + |g(\partial_{\tilde x^i}, \partial_{\tilde x^j}) - g(\partial_{\tilde y^i}, \partial_{\tilde y^j})| = o(|x|^{- \tau}) = o(|\tilde x|^{ - \tau}),\\
&y^i = A_j{}^i\,\tilde y^j + c^i,
\end{align*}
where $A_j{}^i$ and $c^i$ are constants. Note that the second relation and that the metric $g$ is asymptotically flat implies that the matrix $A = (A_j{}^i)$ is orthonormal. Now the argument above show that in defining $m_k(\Phi)$ we can use the coordinate functions $y^i$'s instead of the $x^i$'s. Likewise $m_k(\tilde \Phi)$ can be computed using $\tilde y^i$'s instead of $\tilde x^i$'s. But the frame $\partial_{y^i}$ and $\partial_{\tilde y^i}$ differ from one another by a rigid rotation: $\partial_{y_i} = A_i{}^j\,\partial_{\tilde y^j}$. The argument proving that $\Lambda_k$ is frame independent applies showing that $\Lambda^1_k(\partial_{y^i}) = \Lambda^1_k(\partial_{\tilde y^i})$. This proves that $m_k = m_k(\Phi)$ is independent of $\Phi$.

We have thus shown that
\begin{theorem}\label{Thm:MassDef}
Let $(M,g)$ be a complete $n$-dimensional Riemannian manifold which is asymptotically flat of order $\tau$, i.e. there is a compact set $K$ and a diffeomorphism $\Phi: M \setminus K \rightarrow \RR^n \setminus B_{R_0}$ for some $R_0 > 0$ such that in such coordinate system the metric $g$ satisfies $g_{ij} = \delta_{ij} + o_3(|x|^{-\tau})$. Assume further that the curvature $\Lambda_k$ belongs to $L^1(M)$. Fix $1 \le k < \frac{n}{2}$. If $\tau > \frac{n - 2k}{k + 1}$, then the mass
\[
m_k = \lim_{R \rightarrow \infty} \int_{S_R} (-1)^k\,\Lambda^1_k
\]
is well-defined and is independent of the asymptotic structure at infinity.
\end{theorem}

To finish this section, we give an example. Fix some $1 \leq k < \frac{n}{2}$. Consider an asymptotically flat manifold where the metric takes the following form at infinity
\[
g_{ij} = \exp\Big(\frac{2m}{r^{\frac{n-2k}{k}}}\Big)\delta_{ij} + o_2(r^{-\frac{n-2k}{k}}).
\]

The connection one-forms and the curvature two-forms are
\begin{align*}
\omega_i{}^j 
	&= \Gamma_{ik}^j\,dx^k
	= -\frac{(n - 2k)}{k}\,\frac{m}{r^{\frac{n}{k}}}(\delta_{ij}\,r\,dr + x^i\,dx^j - x^j\,dx^i) + o(r^{1-\frac{n}{k}}),\\
\Omega_i{}^j
	&= d\omega_i{}^j - \omega_i{}^t \wedge \omega_t{}^j \\
	&= \frac{(n - 2k)n}{k^2}\,\frac{m}{r^{\frac{n + k}{k}}}\,dr \wedge (x^i\,dx^j - x^j\,dx^i) -\frac{2(n - 2k)}{k}\,\frac{m}{r^{\frac{n}{k}}} \,dx^i\wedge dx^j  + o(r^{-\frac{n}{k}}).
\end{align*}
Thus,
\[
\Lambda_k^1 \rfloor S_R = c(n,k)\,\frac{(-1)^k\,m^k}{R^n} \sum_{1 \leq i, j \leq n} (x^idx^j - x^j dx^i) \wedge *(dx^i \wedge dx^j)
\]
and so
\[
m_k = c(n,k)\,m^k.
\]
where $c(n,k)$ is some positive constant. 

The above computation also shows that, for the time symmetric slice of the Schwarzschild spacetime in higher dimensions, only the first mass (which is the same as the ADM mass) is nonzero. All the higher order masses, if well-defined, vanish.

\section{On the non-negativity of the $k$-th mass}\label{Sec:PosMass}

It is of interest to see if the $k$-th mass is non-negative under some assumption on either the $\Lambda_k$ or the $\sigma_k$ curvature. We are only able to do so under a very restrictive hypothesis that $(M,g)$ is locally conformally flat and that $\lambda(A_g)$ is asymptotically on the boundary of the $\Gamma_k$ cone. Of the two assumptions, we believe the local conformal flatness assumption is more severe. Note also that, in Schoen and Yau's proof of the positive mass theorem \cite{SchoenYau79-CMP,SchoenYau81-CMP}, one can assume without loss of generality that the manifold is asymptotically scalar flat (i.e. $\lambda(A_g) \in \partial \Gamma_1$).

\begin{theorem}\label{PosMass}
Let $(M,g)$ be a complete, asymptotically flat Riemannian manifold of dimension $n \geq 3$ and let $2 \leq k < \frac{n}{2}$. If, near a given end, $g$ is locally conformally flat, $A_g$ belongs to the $\bar\Gamma_k$ cone and the $\Lambda_k$ curvature vanishes, then the $k$-th mass of that end is non-negative. Furthermore, if the $k$-th mass is zero, then, near that end, $(M,g)$ is isometric to an Euclidean end.
\end{theorem}

The proof of this theorem has a different flavor from what is presented in this paper and will be published elsewhere.


\newcommand{\noopsort}[1]{}
\providecommand{\bysame}{\leavevmode\hbox to3em{\hrulefill}\thinspace}
\providecommand{\MR}{\relax\ifhmode\unskip\space\fi MR }
\providecommand{\MRhref}[2]{%
  \href{http://www.ams.org/mathscinet-getitem?mr=#1}{#2}
}
\providecommand{\href}[2]{#2}

\end{document}